\documentclass[twocolumn,pre,nofootinbib,floatfix,superscriptaddress,showkeys]{revtex4-1}

\usepackage[usenames,dvipsnames,svgnames,table]{xcolor}
\usepackage{tabularx,booktabs,dcolumn}
\usepackage{todonotes,ltxtable}
\usepackage{longtable,array,ragged2e}
\newcolumntype{P}[1]{>{\RaggedRight\arraybackslash\hspace{0pt}}p{#1}}
  
\begin{document}
\title{Network assembly of scientific communities of varying size and specificity}

\author{Daniel T. Citron}
\email{dtc65@cornell.edu}
\affiliation{Department of Physics, Cornell University, Ithaca, New York, 14853 USA}

\author{Samuel F. Way}
\email{samuel.way@colorado.edu}
\affiliation{Department of Computer Science, University of Colorado, Boulder CO, 80309 USA}

\begin{abstract}
How does the collaboration network of researchers coalesce around a scientific topic? What sort of social restructuring occurs as a new field develops? Previous empirical explorations of these questions have examined the evolution of co-authorship networks associated with several fields of science, each noting a characteristic shift in network structure as fields develop. Historically, however, such studies have tended to rely on manually annotated datasets and therefore only consider a handful of disciplines, calling into question the universality of the observed structural signature.
  To overcome this limitation and test the robustness of this phenomenon, we use a comprehensive dataset of over 189,000 scientific articles and develop a framework for partitioning articles and their authors into coherent, semantically-related groups representing scientific fields of varying size and specificity. We then use the resulting population of fields to study the structure of evolving co-authorship networks. Consistent with earlier findings, we observe a global topological transition as the co-authorship networks coalesce from a disjointed aggregate into a dense giant connected component that dominates the network.
  We validate these results using a separate, complimentary corpus of scientific articles, and, overall, we find that the previously reported characteristic structural evolution of a scientific field's associated co-authorship network is robust across a large number of scientific fields of varying size, scope, and specificity.
  Additionally, the framework developed in this study may be used in other scientometric contexts in order to extend studies to compare across a larger range of scientific disciplines.
\end{abstract}


\maketitle

  \section{Introduction}
  A co-authorship network outlines the professional connections between scientific researchers and their collaborators. Co-authorship networks are important objects of study, as they are a measurable representation of the communities that assemble in order to work in an particular area of research. Such communities allow for the transfer of knowledge and skills and sharing of resources required for researching complex problems \cite{Borner49cm24, de1986little, guimera2005team, kaiser2005drawing}.  The assembly of co-authorship networks represents one aspect of the more general problem of understanding the process through which social or collaborative networks attract new members and evolve structurally over time \cite{backstrom2006group, jacobs2015assembling}.

  The recent availability of electronic publishing and online repositories of scientific articles has enabled large-scale studies of scientific research practices \cite{Borner:2004uu, Ginsparg:2004kv, Tabah:1999ve}. 
  In particular, these repositories provide record of collaborations between the authors of each paper, making it possible to construct comprehensive co-authorship networks and analyze their assembly over time. 

  Two recent studies have investigated the development of a small group of research fields (9 and 12 fields, respectively), by measuring the assembly of each field's co-authorship network using a large electronic collection of articles \cite{Bettencourt:2015wh, Bettencourt:2009dl}.  Expanding upon historiographical surveys, they search for patterns in the growth and development of co-authorship networks across different scientific fields.  These studies argue that while each field differs in size and publishing practices (differing in rate of publication, size of collaborations, etc.), nevertheless there appear to be common patterns in how each field's co-authorship network develops.  Specifically, each co-authorship network undergoes a topological transition in which a densely connected giant component of researchers forms over time. This dramatic structural change has been compared to the emergence of a giant component seen in a percolation transition \cite{newman2010networks}, and serves as an empirical indication that the research community undergoes large-scale social reorganization as more researchers join and collaborate with others \cite{Bettencourt:2015wh, Bettencourt:2009dl, guimera2005team}.

  Another study \cite{lee2010complete} takes three example fields (complex networks research; ADS/CFT; Randall-Sundrum model) and describes three stages of development characteristic to co-authorship network assembly in science.  Each network begins as a set of disconnected groups, which then join together to form a large treelike component.  As the research community grows and mixes further, the large component becomes densely connected to itself through the formation of long-range ties.
  This general pattern is consistent with what was reported in \cite{Bettencourt:2015wh, Bettencourt:2009dl}, which also emphasized how the long-range ties between authors created a densely connected community with very short distances between different authors.

  Together, these previous studies suggest the existence of common patterns in how scientific communities assemble over time.  However, they rely on manual annotation of their data, which requires a great deal of labor in order to assemble a co-authorship network.  This in turn limits the number of examples studied and reported on, making it difficult to justify the claim that the patterns observed for a few examples are universal across all scientific fields.  

  In the present study, we propose a framework for analyzing a large population of example topics in order to verify that the development of co-authorship networks, as characterized by earlier studies, is robust across many scientific fields.
  Specifically, we use techniques from natural language processing and machine learning to generate a larger set of example co-authorship networks from the arXiv, a large scientific corpus. We use topic modeling to cluster articles together based on their semantic content, and interpret the clusters of articles as representing different fields of science. We measure the algorithmically-generated co-authorship networks to determine whether they develop in a manner similar to the manually-annotated co-authorship networks studied previously.  We aim to facilitate a larger survey of co-authorship networks across scientific fields first by testing the efficacy of topic modeling as a way to rapidly detect a large number of fields, and then by comparing the assembly behavior of each field's co-authorship network for the purposes of testing whether their growth patterns remain consistent for a large set of fields of varying size and specificity.

  \section{Data Set}\label{data_set}
  The arXiv is an open-access repository of scientific preprints accessible online at \url{www.arxiv.org}.  The site was founded in 1991 and, as of the end of 2016, hosts over 1.1 million articles, primarily in the areas of Physics, Mathematics, and Computer Science \cite{arXivstats16}.  
  Here, we take as our data set the 189,000 articles categorized as Condensed Matter Physics (``cond-mat'' on the arXiv) by the submitting author (or by the arXiv's administrators) during the period starting in April of 1992 and ending in June 2015.  

  The arXiv data have several important advantages for the purposes of the present study.  The articles' full texts and relevant metadata are available to the public. Additionally, arXiv has been well studied from a scientometric perspective \cite{Lariviere:2014}, and has been used to test techniques for algorithmically categorizing scientific articles according to their content \cite{Ginsparg:2004kv}. 

  The set of arXiv articles is only a sample of all published works, and, due to differences in the site's adoption across communities, arXiv's coverage varies from one subfield to the next.  We therefore test that our results obtained by measuring the arXiv actually represent real-world co-authorship networks and not an artifact of the arXiv's incompleteness.  Specifically, to validate our results, we also analyze a subset of the condensed matter articles found on the Web of Science (WoS).  WoS is a database of scientific articles maintained by Clarivate Analytics. We use the 660,000 articles classified as Condensed Matter Physics published between April 1992 and June 2015, requiring that all have titles, abstracts, and author names available in the database \cite{wos2016}. The set of articles from Web of Science partially overlaps with the arXiv data set and represents a complementary data set with non-uniform coverage of the subfields contained on arXiv \cite{Lariviere:2014}.  Using the WoS as a secondary data set makes it possible to verify whether the arXiv contains a truly representative sample of Condensed Matter Physics articles, as well as to check whether the results obtained using the articles from the arXiv are not merely an artifact of the arXiv's incomplete coverage of certain scientific subfields.

  To track the contributions of individual authors, we adopt the convention of labeling each author with their uppercase full names as reported in the publication metadata.  In the context of co-authorship network measurement, this author naming convention errs on the side of splitting individual authors into multiple entities.  That is to say, authors who inconsistently report their names in publications will be counted as multiple separate nodes for the purposes of this study.  This convention also decreases the possibility of many different entities becoming combined into a single composite node, which would artificially collapse together many different nodes in our co-authorship networks.  We verify that our results are robust to changing the author labeling convention by repeating all subsequent analysis using ``[First Initial] [Last Name]'' in Appendix \ref{appendix3}.  Larger-scale analyses involving a broader reach of disciplines will require additional steps to disambiguate author identities (such as the tools described in \cite{bhattacharya2007collective, song2007efficient}).  After preprocessing author names in this way, the arXiv data set includes 96,000 unique authors.

  For the purposes of text mining and topic modeling we focus on each article's title and abstract under the assumption that authors write titles and abstracts with the intention of concisely summarizing an article's contents. Past studies have argued that focusing on article abstracts has the additional benefit of minimizing the amount of ``structural'' text processed by the topic model, allowing the inferred topic structures to focus on field-specific content, rather than commonalities in presentation of the English language \cite{Ginsparg:2004kv,joachims:svms}.

  \section{Methods}\label{methods}

  \subsection{Topic Model}

  Past studies exploring the formation of co-authorship networks have relied on manual annotation to determine which authors contribute to and are therefore considered part of a scientific field \cite{Bettencourt:2015wh, Bettencourt:2009dl, lee2010complete}. 
  This approach, however, requires a great deal of human effort and, consequently, has been applied to only a few disciplines and with somewhat arbitrary definitions of which publications and authors belong to the community in question. It therefore remains unclear how robust past results are to varying the criteria for selecting communities, and for varying levels of specificity governing the breadth and size of such communities. 

  To address these limitations, we introduce an approach that uses topic modeling to automate the process of identifying groups of semantically-related documents and partitioning their authors into fields corresponding to their areas of expertise \cite{Boyack:2005vv}. As a consequence of the number of documents belonging to a given subfield and the commonality of its language, the topics and thus the fields extracted by this technique will vary in terms of size and specificity, yielding a population of corresponding co-authorship networks. That is, we can test whether the reported structural patterns are robust to varying definitions of sub-community. At the same time, we explore the usefulness of topic modeling as an automated, scalable means for partitioning the global network of all researchers into co-authorship networks organized around specific fields.

  Topic modeling is an unsupervised machine learning technique that characterizes the underlying thematic content of a given corpus by identifying groups of semantically-related, co-occurring words---the ``topics''---while simultaneously identifying the proportion of each topic present in each document in the corpus. Here, we use latent Dirichlet allocation (LDA) \cite{blei2003latent, griffiths2004finding}, a popular topic model that produces static definitions for topics, formalized as probability distributions over all words in a given vocabulary. Accordingly, for each document the model infers a distribution over these topics.

  Prior to applying topic modeling, we utilize several common natural language processing techniques to preprocess the corpus text. In particular, we combine the text from each article's title and abstract into a single document, remove all non-alphabetic characters, and convert all letters to lowercase. Common English stop words (``the,'' ``and,'' ``of,'' etc.) are also removed, as well as certain words that appear very commonly in the arXiv data set but that contain no scientific content (numbers, names of publishers, ``thank you,'' etc.). The document text is also lemmatized in order to increase the likelihood of discovering overlaps in the word usage within and between documents

  After preprocessing all articles, we use MALLET \cite{McCallumMALLET}
  , an open-source implementation of LDA, to train a series of topic models, varying the number of topics between $k\!=\!25$ and $k\!=\!100$. As expected, for small $k$, LDA produces broadly-defined topics, and for large $k$, more narrowly-defined topics. For our purposes, $k\!=\!50$ provides sufficient resolution for the model to recover topics that resemble established subfields within condensed matter physics.  We emphasize that we do not intend to use this topic model to represent the optimal or definitive partition of arXiv according to subject matter.  Rather, our model provides a large set of readily-interpretable topics, varying in both size and specificity, allowing us to test the robustness of past claims against a heterogeneous population of fields and their corresponding authors.
  We present our analysis of the $k\!=\!50$ topic model below and note that our results are robust to small changes in $k$. That is, the results that we report below do not change significantly if we repeat our subsequent analyses using a model with $k\!=\!45$ or $k\!=\!55$ topics.

  After training our topic model, we manually inspect each topic to determine whether it resembles a field of condensed matter physics.
  As an example, the most probable words associated with Topic 5 include keywords such as ``quantum,'' ``state,'' ``qubit,''  ``entanglement,'' and ``decoherence.''  Looking at the set of articles to which the topic model assigns a high probability ($P(\text{Topic}=5) > 0.6$), we find articles such as  ``Demonstration of Two-Qubit Algorithms with a Superconducting Quantum Processor'' (0903.2030) and ``Controllable coupling between flux qubits'' (cond-mat/0507496).  Together, these observations suggest that articles strongly associated with Topic 5 are related to quantum computing and quantum information. We also check that the articles identified by the topic model do not merely reflect clusters of articles specific to arXiv by inferring topics on the articles belonging to the Web of Science (WoS) data set.  In the case of Topic 5, we find articles such as ``Flexible two-qubit controlled phase gate in a hybrid solid-state system'' and ``Two-electron coherence and its measurement in electron quantum optics,'' which confirms that articles associated with Topic 5 appear to be related to quantum computing.

  In addition to quantum computing, LDA recovers topics resembling other established subfields of condensed matter physics, including spin glasses (Topic 1); Bose-Einstein condensates (Topic 3); magnetic materials (Topic 19); glassy physics (Topic 28); topological phases (Topic 30); and cuprate superconductors (Topic 43).  (Refer to Appendix \ref{appendix1} to see each topic's interpretation.)

  \subsection{Co-authorship Network Generation}

  \begin{figure*}[ht!!]
      \centering
      \vspace*{2mm}
  \includegraphics[width=.9\textwidth]{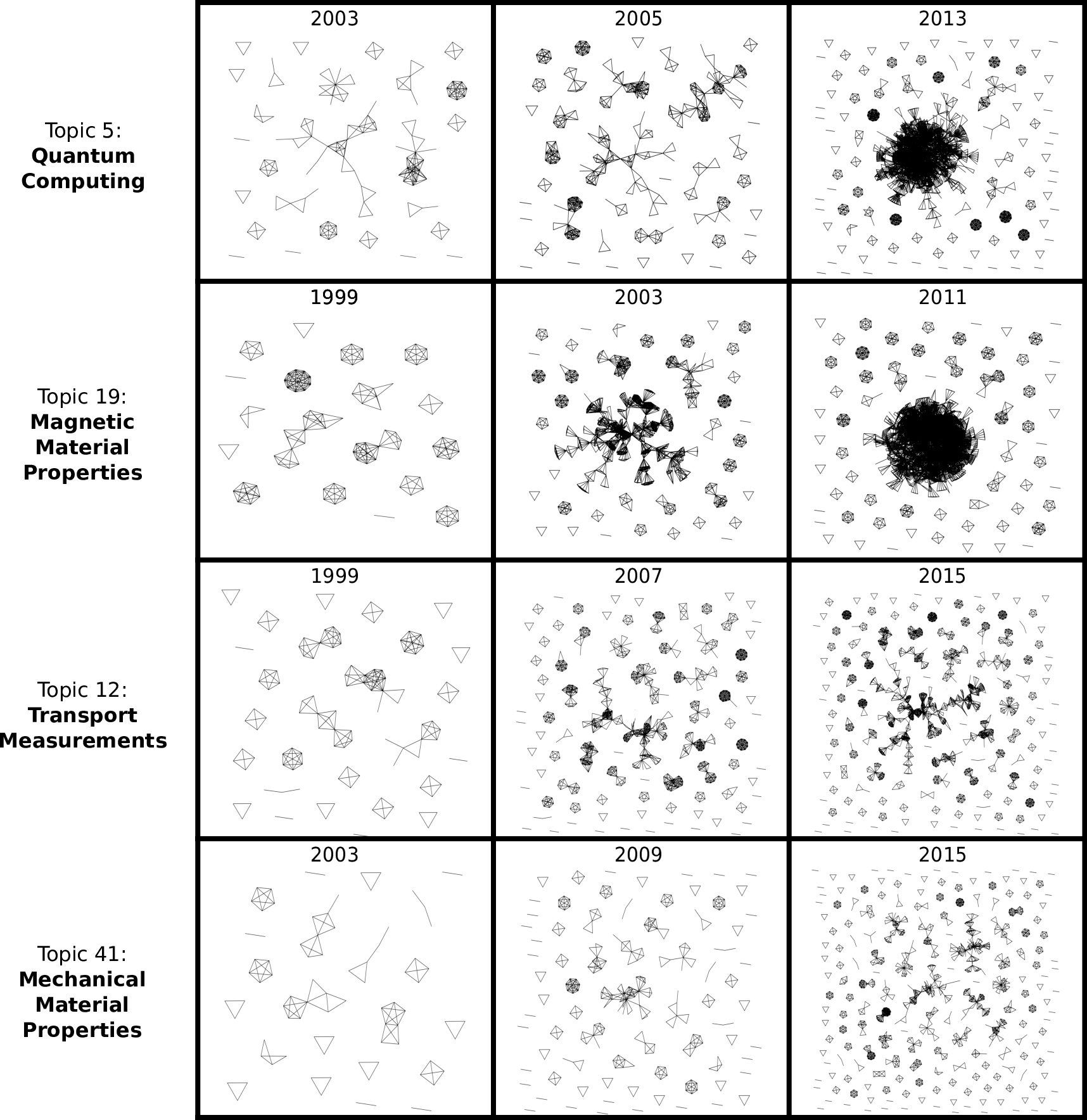}
  \caption{Examples of different network behaviors.  Each row shows how a topic's co-authorship network develops over time, with network snapshots labeled by the year observed.  Each node represents an author, and each edge represents a collaboration between the two authors.  Disconnected single nodes are not pictured.  The top two rows (Topics 5 and 19) illustrate three stages of network assembly: disjointed group of cliques; treelike connected cluster of cliques; densely connected giant component.  The third row (Topic 12) is an example of a network that only forms a treelike largest component.  The bottom row (Topic 41) is an example of a network that forms no single giant component.}\label{vis}
  \end{figure*} 

  We use our topic model to construct a set of co-authorship networks, where each network represents the set of authors that produced the articles strongly associated with one of the topics discovered by the topic model. We emphasize that the topic modeling algorithm is only given information related to the textual content of the articles and receives no information about authorship, authors' collaborative relationships, or publication dates.  While there are topic modeling algorithms that do take into account other links between documents (e.g. \cite{guo2009latent, rosen2004author}), we want to determine whether textual content is sufficient to reproduce patterns in how groups of researchers in the same related form a collaborative community.

  We find the articles that are primarily associated with each topic $t$ by selecting the subset of articles assigned a probability weight $P(t) > 0.6$.  The cutoff at $0.6$ selects articles that are strongly associated with one particular topic, but is not so strict that it excludes too many articles.   With $P(t) > 0.6$, we associate between 100 and 3000 arXiv articles with each topic. We also use an alternative thresholding criterion to check whether the choice of thresholding biases our results.  We repeat all subsequent analysis using a second method of categorizing articles whereby each article is assigned to the smallest set of topics that account for 50\% of its subject matter. All reported results are robust to varying the thresholding scheme.

  We construct a co-authorship network by identifying the authors of each topic's associated articles.  Each author is represented in the topic's network as a node.  Two author nodes are linked by an edge if they have written an article together. \cite{Newman16012001,Newman:2004fw}.  Hence, a group of authors who collaborated on an article together appears in the network as a fully connected clique, and two articles with multiple authors in common will appear in the network as overlapping cliques that share nodes.  (We also use a modularity score to measure the extent to which authors associated with different topics connect to one another.  We find that our topic model does tend to sort authors into distinct communities in \ref{appendix4})

  We reconstruct each co-authorship network's assembly and growth over time using each month of arXiv's operation from April 1992 through June 2015 as a discrete time step. At each time step we include in the network all author nodes that have written articles at or prior to the current time step.  We also connect all pairs of author nodes that have collaborated on one or more articles at or prior to the current time step.

  \section{Results}\label{results}

  \subsection{Co-Authorship Network Measurements}

  \begin{figure*}[ht!!]
      \centering
      \vspace*{2mm}
  \includegraphics[width=1.\textwidth]{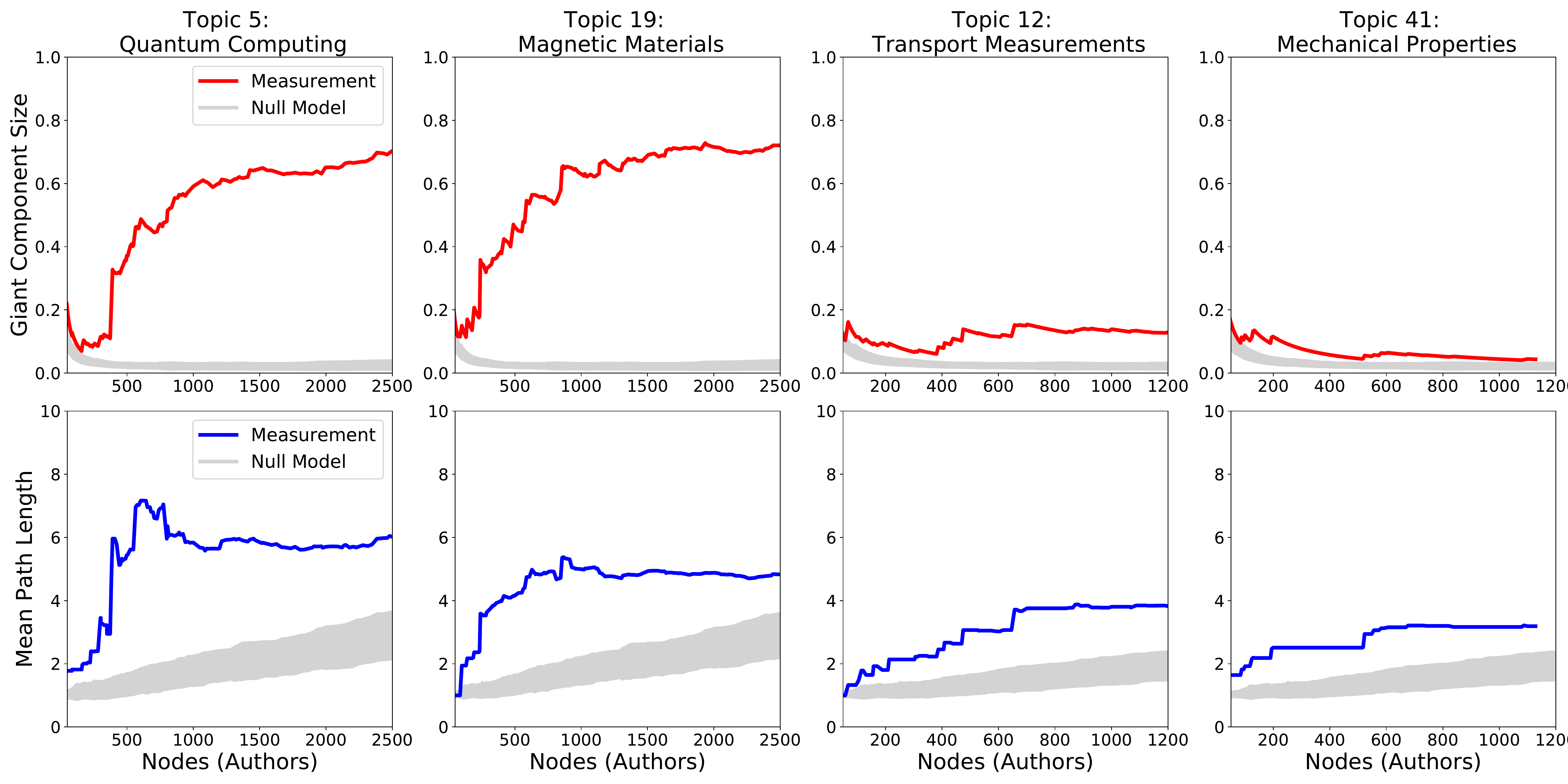}
  \caption{Quantitative measurements of co-authorship networks.  The top row shows the fraction of nodes belonging to the largest component as a measure of network size, plotted vs. the total number of nodes in the network.  The bottom row shows the mean geodesic path length of the largest component (``mean path length'') vs. the total number of nodes in the network.  
  For Topics 5 and 19, the largest component grows to dominate the network.  As the largest component grows, its mean path length increases quickly at first and then begins to decrease.
  For Topic 12, a single large component grows, but remains treelike and its mean path length only continues to increase.
  For Topic 41, no giant component forms. } \label{measurements}
  \end{figure*}

  Figure \ref{vis} shows the network growth for four different example topics: quantum computing (Topic 5), magnetic material properties (Topic 19), transport measurements (Topic 12), and mechanical properties of materials (Topic 41).
  For the first two topics in Figure \ref{vis} there appear to be three separate stages through which the giant component develops.  Each network begins as a disjointed set of cliques, as the authors who share a field publish in separate groups.  Next, a few of the cliques join together, forming a loosely connected, almost tree-like backbone of connected cliques.  
  In the final stage, enough cliques overlap with one another such that the largest connected component becomes densely connected. This characteristic three-stage pattern is consistent with what has been reported previously \cite{lee2010complete}. 
  By contrast, the largest component of Topic 12's network only grows to reach the treelike stage, and Topic 41's network has no giant component.

  We confirm this interpretation of the network visualizations by measuring various properties of each topic's co-authorship network.  We measure the fraction of nodes belonging to the largest connected component (``giant component size'').  We also measure the giant component's mean geodesic path length between all pairs of nodes belonging to the giant connected component (``mean path length'').  The mean path length ranges between a minimum for fully connected networks and a maximum for treelike networks, and so serves as a measure of how closely connected the individuals belonging to the giant component are to one another \cite{Bettencourt:2009dl, leskovec2005graphs, newman2010networks} 

  Figure \ref{measurements} shows measurements of the size and mean path length of the giant component for each of the topics shown in Figure \ref{vis}.  For Topics 5 and 19 (two leftmost columns), the giant component's size increases steadily as more and more nodes are added to the network.  At the same time, the mean path length first increases as the giant component grows initially and then peaks and decreases \cite{leskovec2005graphs}.  This non-monotonic behavior suggests two stages in the development of the giant component: initial growth as cliques first start to overlap with one another, and densification when enough ``long-range'' edges form to reduce the average distance between authors \cite{lee2010complete, newman2010networks, Watts:1998db}  These two growth stages are consistent with a treelike cluster of cliques that becomes a densely connected cluster. 
  As a point of comparison, the largest component in Topic 12 does grow to include a large fraction of the nodes in the network, but its mean path length increases steadily over time.

  The co-authorship network development patterns are not merely the result of sampling a large number of articles that join together by chance. For comparison, we consider a null model in which articles are grouped together at random, rather than grouped together according to topic modeling, to test whether the topic modeling is responsible for identifying the clusters of authors.  For each instance of the null model, thousands of articles are selected from the arXiv cond-mat data set at random.  The co-authorship network of this randomly-selected group of articles is then constructed, and the properties of the largest connected component are measured.  The results of this null model are plotted in gray in Figure \ref{measurements}, where the vertical height of the gray region represents the mean $\pm$ one standard deviation across 100 instances of the null model.  The null model's average behavior contrasts dramatically with the measurements of the scientific co-authorship networks identified using the topic model.
  These results strongly suggest that the aggregation of authors to form a giant, densely connected component is not merely the result of sampling an arbitrary subset of arXiv.  Rather, it appears that the topic model, which was given no information about authorship or other such links between documents, was able to identify clusters of researchers based on their textual content alone.  The nonrandom grouping of authors further validates the topic model's meaningful clustering of articles: the articles represent the output of an association of researchers with similar interests.

  The example topics shown in Figure \ref{vis} and Figure \ref{measurements} exemplify three general types of network assembly observed for the other topics.  
  Out of the 50 topics, 22 have co-authorship networks that undergo the transition from a scattered collection of cliques; to an extended, treelike connected group of cliques; to a densely connected giant component.  These results are qualitatively consistent with those obtained earlier for groups of articles annotated by human experts \cite{Bettencourt:2015wh, Bettencourt:2009dl}. 
  From the remaining topics, 17 form a single large component that occupies a small fraction of nodes in the network, but have not yet formed enough long-range ties that the network mean path length stops growing monotonically. 
  The remaining 9 topics show little or no sign that they form any giant connected component.  (Refer to Appendix \ref{appendix2} for a summary of all co-authorship networks' behavior.)

  Finding that a topic's corresponding co-authorship network does not form a densely connected GCC does not necessarily suggest that the research field is not well-established.  There are several possible reasons why a dense giant component does not form in all cases.
  The existence of a giant component only indicates that there are a great many researchers that have collaborated with one another.  Inter-group collaborations may be more frequent in some fields than in others, and a giant component is only likely to form when there are many collaborations between research groups.
  Additionally, the arXiv does not necessarily represent a comprehensive sampling of articles from all subfields of science. The arXiv's coverage of some fields may be incomplete, such as microscopy (Topic 15) and surface chemistry (Topic 47).

  \subsection{Validation Across Corpora}

  \begin{figure*}[ht!!]
      \centering
      \vspace*{2mm}
  \includegraphics[width=.95\textwidth]{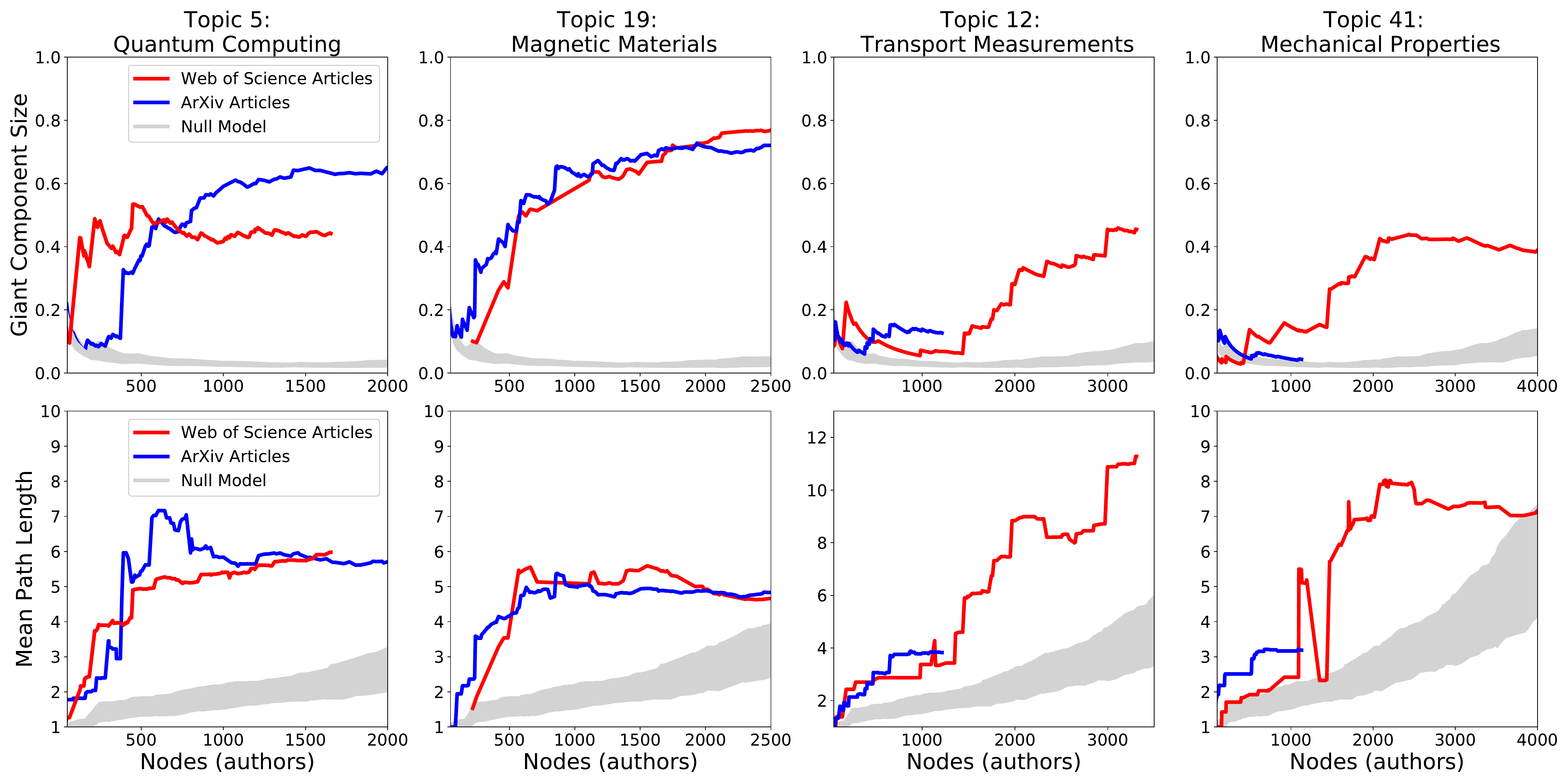}
  \caption{Comparison between co-authorship networks generated from arXiv and Web of Science.  Each column corresponds to a different topic.  The top row shows the fraction of nodes belonging to the largest component as a measure of network size vs. the total number of nodes in the network.  The bottom row shows the mean geodesic path length of the largest component (``mean path length'') vs. the total number of nodes in the network. Each plot shows the measurements made of the co-authorship network from the Web of Science (in red), from  arXiv (in blue), as well as co-authorship networks generated from randomly chosen articles from Web of Science (null model, in gray). }\label{wos_comp} 
  \end{figure*}

  The characteristic growth patterns seen for the co-authorship networks of authors from the arXiv remain consistent when we repeat the same analysis using another corpus. We use the topic model trained on the arXiv data to infer topics for the condensed matter physics articles from the Web of Science (WoS).  The same procedures for generating and measuring the co-authorship networks for the WoS articles reveals that the topic model trained on the arXiv is still able to identify large connected clusters of articles in the WoS. Figure \ref{wos_comp} compares the behavior of the co-authorship networks that occur within both the arXiv and WoS. 

  In the majority of cases, the co-authorship networks identified from the WoS articles behave similarly to the ones identified on arXiv.  For example, the co-authorship networks for research on quantum computing and magnetic material properties (Topics 5 and 19, the two leftmost columns of Figure \ref{wos_comp}) form a dense giant component for both arXiv and for WoS.
  There is also a group of topics whose networks form only a treelike giant component or no giant component in the arXiv data but do form a dense component with a shrinking mean path length in the WoS data.  Topics that do this include transport measurements and mechanical material properties (Topics 12 and 41, shown in the two rightmost columns of Figure \ref{wos_comp}), as well as nanoscale devices (Topic 16) and inelastic scattering experiments (Topic 33).  We note that these topics have an experimental focus.  Experimental research subjects are known to have less coverage on arXiv, but are covered more comprehensively in the WoS \cite{Lariviere:2014}.  There are also a few topics with decreased coverage on WoS because the WoS does not categorize them as condensed matter.  For example, articles on ultracold atoms (Topics 3 and 20) may be categorized separately as ``atomic, molecular, and optical physics'' and articles on soft condensed matter (Topics 25 and 50) may be categorized separately as ``fluids.''  Consequently, these topics' decreased inclusion in the WoS data set leads to smaller and less densely connected co-authorship networks.

  Overall,27 out of 50 topics have co-authorship networks that develop similarly for both the WoS data and the arXiv data (Appendix \ref{appendix2}).  Additionally, 10 experimentally-focused topics have co-authorship networks that grow to form large giant components on on account of having better coverage on the WoS compared to the arXiv.  Another three topics (Topics 9, 10, and 42) have very low coverage on the arXiv (fewer than 100 associated articles) and do not form giant connected components with either the arXiv or the WoS.  Given that, across both corpora, none of these three topics has many strongly associated articles, it is likely that Topics 9, 10, and 42 are actually ``junk topics,'' meaning that they do not reflect coherent themes and so are not useful for the purposes of the present study. The consistency of the behavior of these co-authorship networks measured across different corpora suggests that the collaborative communities identified using the model are reflected in multiple data sets.

  \subsection{Robustness to Edge Removal}

  \begin{figure*}[ht!]
      \centering
      \vspace*{2mm}
  \includegraphics[width=1\textwidth]{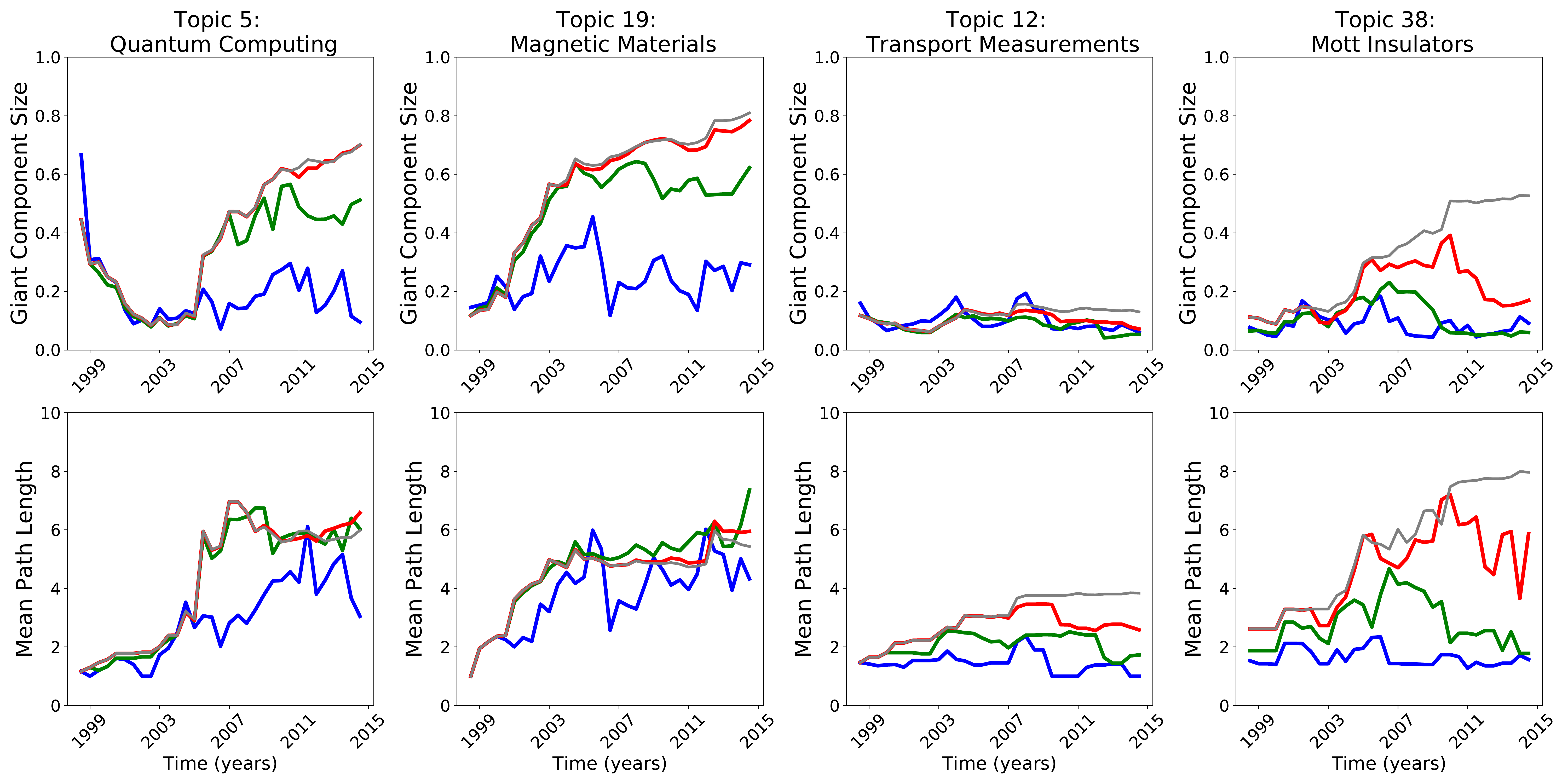}
  \caption{Network Robustness to Edge Removal. For four topics, we show how the network assembly changes when edges only remain in the network for a limited amount of time.  Each plot shows the network's giant component size over time for four different edge lifetimes.  For short edge lifetimes (2 years in blue; 5 years in green), the giant connected component fails to develop or develops much more slowly compared to the permanent edge (``no limit,'' gray) case.  For longer edge lifetimes (10 years, red), the giant component approaches the no limit case.}\label{lifetime}
  \end{figure*}

  Finally, we address the question of whether the co-authorship network development patterns seen in our data and in previous studies are robust to relaxing the assumption that all edges in the co-authorship network are maintained indefinitely after they are established.  
  Previous studies have constructed co-authorship networks wherein that collaborative link, once established, are maintained forever \cite{Bettencourt:2015wh, Bettencourt:2009dl, lee2010complete}. In practice, when such a collaborative relationship requires significant efforts to maintain, this assumption is not necessarily valid.

  We re-assemble the co-authorship networks for each of the topics, this time allowing edges to expire after a fixed number of months.  That is to say, if two authors do not repeat a collaboration after a certain amount of time, the edge representing their relationship is removed from the network.  The results are plotted in Figure \ref{lifetime}, where the uppermost curve (gray; ``no limit'') shows how the giant component grows if edges survive indefinitely, while the other curves show how those measurements change if the edges are removed after 2 (blue), 5 (green), or 10 (red) years.

  Limiting the lifetime of edges to a few years causes giant components to develop much more slowly, or to not develop at all.  For Topics 5 and 19, the network measurements for 5 and 10 years are very close to the indefinite lifetime limit.  This suggests that these networks are particularly robust to edge removal, reflecting a very densely connected giant component where edges are frequently renewed \cite{lee2010complete}.  For Topic 38, the giant component forms much more slowly, and actually begins to disassemble for edge lifetimes of 2 or 5 years.  For Topic 12, finite edge lifetime only suppresses the component formation of a large component.  (Appendix \ref{appendix5} contains additional visualizations of these graphs, comparable to those appearing in Figure \ref{vis}.)

  Currently, it is unknown what criteria for including and excluding nodes and edges from co-authorship network models best reflect the reality of authors entering and exiting different fields.  What is clear, however, is that the assumption that the relationships represented by edges between authors last forever is important for obtaining the quantitative results that reflect a topological transition in the co-authorship network.  Shortening the lifetime of edges can dramatically change a co-authorship network's evolution over time.

  \section{Discussion}\label{discussion}
  This study expands upon previous research exploring the growth and development of co-authorship networks using topic modeling to algorithmically identify and study a large population of scientific fields, along with their associated articles and authors.  Our results show that, for the topics determined using LDA, a large majority of co-authorship networks undergo a topological transition to form a densely-connected giant component characterized by three stages of development.  These patterns corroborate findings from earlier studies that focused on small numbers of (often manually assembled) co-authorship networks. Our results demonstrate that the characteristic topological transition is robust to variations in the definition of a scientific field, both in terms of size and specificity. Additionally, our methods employ algorithmic clustering and require no input from human experts, yet the results are largely consistent with previous studies.  
  We also found that the patterns in co-authorship network development are consistent across corpora, which we demonstrate by repeating our analysis using data from both the arXiv and the Web of Science.  One notable difference between the two corpora is reflected in how arXiv's selections of articles related to certain experimentally-focused topics are under-populated: in these cases, the co-authorship networks constructed using the larger WoS data set undergo a topological transition, while the corresponding networks drawn from the arXiv data do not. 
  
  Topic modeling is a rich and actively growing area of research within the statistical modeling and natural language processing communities. In our study, we used latent Dirichlet allocation, one of the most popular yet simplest forms of topic modeling. This model assumes a static definition for topics and thus scientific communities, which are known evolve with time. Additionally, the model does not directly incorporate other, non-semantic relationships between documents (such as co-authorship or citations), which may signal alternate forms of cohesion within a scientific community. For our purposes, we consider the assembly and development co-authorship networks over relatively short periods of time and thus favor LDA's straightforward approach.
  Future work in this area, however, should explore more sophisticated algorithms that consider topic dynamics (e.g. \cite{blei2006dynamic, Wang:2006vw}) and additional measures of community cohesion in order to more thoroughly address the co-evolution of scientific fields.

  Our method for algorithmically generating and analyzing a large number of fields can also be used as a framework for further exploring the claims made in a wide variety of bibliometric contexts.  For example, one could also perform a comparison of the micro-scale dynamics of individual authors many different fields.  Recent studies have used agent-based models of author behavior to explain the patterns in publishing behavior that one sees in different fields of science (e.g. \cite{Boyack:2005vv,sun2013social}).
  Once again, most of these studies have relied on manually annotated data sets, and as such, they have historically been limited to only a handful of fields. The approach that we develop in this study, however, enables future work, in conjunction with comprehensive data sets like the arXiv or Web of Science, to further test the accuracy of these models of author behavior across a large and diverse population of scientific fields.

  \section{Acknowledgments}
  This material is based upon work supported by the National Science Foundation Graduate Research Fellowship under Grant No.\ DGE-1144153, and NSF award SMA 1633747. Any opinion, findings, and conclusions or recommendations expressed in this material are those of the authors and do not necessarily reflect the views of the National Science Foundation. The authors would also like to acknowledge Michael W.\ Macy, Paul H.\ Ginsparg, Alexandra Schofield, and Haofei Wei, as well as Brent Schneeman, Laurence Brandenberger, Richard Barnes, and the other attendees of the Santa Fe Institute's 2015 Complex Systems Summer School for helpful discussions.

%


\begin{onecolumngrid}
\bigskip

\setcounter{equation}{0}
\setcounter{section}{0}
\setcounter{subsection}{0}
\setcounter{figure}{0}
\setcounter{table}{0}
\setcounter{page}{1}
\makeatletter
\renewcommand{\theequation}{S\arabic{equation}}
\renewcommand{\thefigure}{S\arabic{figure}}

\vspace{-.5cm}
\begin{appendix}
  \section{Interpreting Topic Model Output}\label{appendix1}
  This table lists the properties of each of the $k=50$ topics identified using LDA trained on the condensed matter physics (cond-mat) articles on arXiv.  \textbf{Keywords} represent a few of the words most strongly associated with a topic.  The \textbf{Example Article} is the reference number of an article with a high probability assigned to a topic.  We interpret each topic as representing a research field, and the \textbf{Interpretation} is the name that we use to refer to that research field.

\begin{center}
\begin{longtable}{| P{.025\textwidth} | P{.5025\textwidth} | P{.15\textwidth} | P{.3\textwidth} |}

\hline \multicolumn{1}{|l|}{\textbf{\#}} & \multicolumn{1}{c|}{\textbf{Keywords}} & \multicolumn{1}{c|}{\textbf{Article}} & \multicolumn{1}{c|}{\textbf{Interpretation}} \\ \hline 
\endfirsthead

\multicolumn{4}{c}%
{{\bfseries \tablename\ \thetable{} -- continued from previous page}} \\
\hline \multicolumn{1}{|l|}{\textbf{\#}} & \multicolumn{1}{c|}{\textbf{Keywords}} & \multicolumn{1}{c|}{\textbf{Article}} & \multicolumn{1}{c|}{\textbf{Interpretation}} \\ \hline 
\endhead

\hline \multicolumn{4}{|r|}{{Continued on next page}} \\ \hline
\endfoot

\hline \hline
\endlastfoot

  1 & critical scaling ising transition temperature spin glass dimension random order phase correlation lattice & cond-mat/9709165 & Spin Glasses; Magnetic Frustration \\
  \hline
  2 & network node distribution degree random graph complex dynamic population scalefree market pattern & cond-mat/9709165 & Complex Networks; Population Dynamics \\
  \hline
  3 & condensate boseeinstein atom trap gas bose interaction potential condensation trapped atomic bec & 0812.0499 & Bose-Einstein Condensates \\
  \hline
  4 & pressure phase alloy compound gpa temperature transition structural crystal lattice diffraction xray superconductivity & 0712.2955 & Superconducting Phases; High Pressure Phases \\
  \hline
  5 & quantum state qubit entanglement spin dot decoherence coupling single control gate coupled information& 0903.2030 & Quantum Computing; Quantum Information \\
  \hline
  6 &  dynamic noise quantum state oscillator dynamical regime frequency nonequilibrium driven coupled evolution fluctuation& 1411.2637 & Quantum Oscillators  \\
  \hline
  7 &  spin magnetic ferromagnetic magnetization effect current anisotropy polarization exchange layer interaction coupling& 1205.2835 & Spins in Materials; Spintronics  \\
  \hline
  8 &  experimental recent theoretical physic experiment phenomenon present review work physical discus understanding & 1306.1774 & Review Articles  \\
  \hline
  9 &  coupling interaction spinorbit phonon electronphonon effect phonons electron strong mode rashba polaron & cond-mat/9911404  & Polarons \\
  \hline
  10 &  phase transition diagram order critical temperature point state quantum region firstorder behavior & cond-mat/0602237  & Phase Transitions; Quantum Phase Transitions \\
  \hline
  11 &  quantum optical dot exciton semiconductor emission electron energy excitons hole laser excitation  & 0906.3260  & Quantum Dots; Mesoscale Physics\\
  \hline
  12 &  temperature conductivity thermal transport dependence low effect resistivity heat coefficient scattering thermoelectric  & cond-mat/0210047 & Transport Measurements \\
  \hline
  13 &  wave soliton nonlinear periodic lattice potential instability velocity oscillation mode dynamic propagation  & 0904.4417 & Solitons; Stationary States \\
  \hline
  14 &  vortex magnetic pinning lattice superfluid flux core superconductors current critical superconducting defect & cond-mat/9908317 & Superconductor Vortices \\
  \hline
  15 & scanning microscopy measurement tunneling image force local tip surface imaging probe atomic resolution & 1009.2393 & Microscopy \\
  \hline
  16 & device material application design control cell efficiency performance memory potential power circuit technology& 0804.1389 & Electronic Devices \\
  \hline
  17 &  approximation density potential energy calculation solution effective functional exact expression expansion order & cond-mat/0007282 & Mathematical Physics \\
  \hline
  18 & spin lattice chain magnetic quantum heisenberg state interaction antiferromagnetic phase order exchange & 1404.0194 & Magnetic Frustration; Spin Chains \& Lattices \\
  \hline
  19 & magnetic temperature heat measurement magnetization susceptibility transition specific crystal single compound ferromagnetic & 1411.2135 & Magnetic Material Properties \\
  \hline
  20 & gas lattice atom interaction fermi superfluid optical boson fermion ultracold state quantum & 0806.4310 &  Ultracold Atoms Dynamics \\
  \hline
  21 & film thin layer substrate temperature sample surface growth thickness grown deposition nanoparticles  & 1502.07223 & Oxide Thin Films \\
  \hline
  22 & spin relaxation magnetic nuclear electron temperature rate resonance nmr dynamic frequency hyperfine  & 1501.02897 & Nuclear Magnetic Resonance \\
  \hline
  23 & quantum hall electron magnetic state effect landau level fractional twodimensional edge filling  & 1109.6219 & Quantum Hall Effect \\
  \hline
  24 & nanotube carbon nanowires transistor device gate channel effect voltage nanowire tube contact transport  & 1112.4397 & Electronic Devices; Nanoscale Devices\\
  \hline
  25 & polymer chain protein interaction dna solution simulation length charge molecule force charged concentration & cond-mat/0504108 & Soft Condensed Matter; Polymer Physics\\
  \hline
  26 & impurity disorder interaction kondo liquid localization effect disordered electron quantum fermi anderson  & 1209.1606 & Disordered Systems\\
  \hline
  27 & frequency optical cavity mode light microwave wave resonance resonator dielectric radiation photonic & 1212.0237 & Optics; Metamaterials \\
  \hline
  28 &  dynamic glass liquid temperature simulation relaxation transition molecular water density fluid correlation glassy & 1209.3401 & Glasses \\
  \hline
  29 & graphene layer edge bilayer electronic dirac gap band monolayer graphite sheet nanoribbons & 1309.5398 & Graphene \\
  \hline
  30 &  topological symmetry state insulator phase quantum fermion gauge dirac chiral majorana breaking edge & cond-mat/0506581 & Topological Phases \\
  \hline
  31 & simulation monte carlo algorithm problem numerical quantum present efficient technique scheme calculation & 0705.4173 & Simulation Methods; Monte Carlo \\
  \hline
  32 & quantum dot transport conductance tunneling electron current effect voltage charge lead contact junction & 0706.2950 & Mesoscale Transport \\
  \hline
  33 & scattering mode spectrum excitation peak frequency energy optical raman neutron inelastic phonon & cond-mat/0308170 & Inelastic Scattering Experiments \\
  \hline
  34 & 	distribution random correlation matrix statistic fluctuation probability gaussian ensemble large statistical density & cond-mat/9704191 & Condensed Matter Theory; Random Matrices \\
  \hline
  35 & flow particle granular fluid velocity shear force dynamic simulation friction hydrodynamic viscosity  & cond-mat/9511105 & Soft Condensed Matter; Granular Physics \\
  \hline
  36 & current junction josephson superconducting ring magnetic flux critical effect array wire temperature tunnel & cond-mat/9811017 & Superconducting Devices; Josephson Junctions \\
  \hline
  37 & entropy equilibrium energy nonequilibrium fluctuation statistical heat thermodynamic distribution relation thermodynamics theorem temperature & 1111.7014 & Thermodynamics \\
  \hline
  38 & hubbard interaction electron correlation charge mott state lattice insulator correlated phase coulomb band hopping & cond-mat/0508385 & Mott-Hubbard Model \\
  \hline
  39 & band surface fermi state electronic gap electron energy photoemission calculation level spectroscopy  & 1101.5615 & Electronic Spectra; ARPES \\
  \hline
  40 & scaling exponent percolation size cluster dimension critical alpha lattice law fractal distribution & cond-mat/0608223 & Critical Phenomena \\
  \hline
  41 & stress elastic strain material deformation dislocation shear modulus mechanical crack solid response fracture & cond-mat/0410642 & Mechanical Properties of Materials \\
  \hline
  42 & state energy ground bound number spectrum density excited level particle potential excitation  & cond-mat/9712133 & Quantum States \\
  \hline
  43 & superconducting superconductivity doping superconductors cuprates temperature order state pseudogap magnetic charge & 1504.06972 & Cuprate Superconductors \\
  \hline
  44 & magnetic ferroelectric phase transition orbital ordering polarization temperature order manganite state charge & 1309.0291 & Ferroelectrics \\
  \hline
  45 & matrix quantum entanglement operator boundary lattice chain entropy exact group solution spin representation & cond-mat/0211081 & Condensed Matter Theory \\
  \hline
  46 & superconducting state superconductors superconductivity superconductor gap pairing symmetry dwave temperature order pair  & cond-mat/0307345 & Superconductivity \\
  \hline
  47 & surface interface domain wall growth boundary force nucleation droplet bulk substrate layer & 0809.1779 & Surface Physics; Surface Chemistry \\
  \hline
  48 & calculation atom energy density molecule electronic surface functional molecular cluster defect hydrogen & 1312.4272 & Density Functional Theory \\
  \hline
  49 & particle diffusion process motion dynamic brownian rate reaction random stochastic transport probability  & 1207.6190 & Nonequlibrium Stat Mech; Stochastic Processes \\
  \hline
  50 & crystal membrane nematic liquid surface curvature defect order rod orientation elastic phase & 1304.0575 & Soft Condensed Matter; Structured Fluids \\
  \hline
\end{longtable}
\end{center}

  \section{Network Assembly Results for All Topics}\label{appendix2}

  This table summarizes the behavior of each topic's corresponding co-authorship network.  For each topic (denoted by \textbf{\#} and \textbf{Interpretation}), we show the number of articles for both the arXiv and Web of Science data sets (\textbf{\# Articles arXiv} and \textbf{\# Articles WoS}, respectively).  Also shown is the assembly behavior of the co-authorship network for each topic (\textbf{GC Transition}).  Referring back to Figures \ref{vis} and \ref{measurements}, ``No GC'' refers to no giant component formation, where cliques of authors remain disjointed.  ``Treelike GC'' refers to cases where cliques of authors join together to form an extended, treelike giant component that has a mean path length that continues to grow. ``Dense GC'' refers to cases where cliques join together to form a densely connected giant component with many overlapping cliques and a mean path length that increases and then decreases.
  
  \begin{center}
\begin{longtable}{| P{.025\textwidth} | P{0.35\textwidth} | P{.14\textwidth} | P{.14\textwidth} | P{.14\textwidth} | P{.14\textwidth} |}

\hline 
\textbf{\#} & 
\textbf{Interpretation} & 
\textbf{\# Articles arXiv} & 
\textbf{GC Transition arXiv} & 
\textbf{\# Articles WoS} & 
\textbf{GC Transition WoS} \\ \hline 
\endfirsthead

\multicolumn{6}{c} 
{{\bfseries \tablename\ \thetable{} -- continued from previous page}} \\ \hline
\textbf{\#} & 
\textbf{Interpretation} & 
\textbf{\# Articles arXiv} & 
\textbf{GC Transition arXiv} & 
\textbf{\# Articles WoS} & 
\textbf{GC Transition WoS} \\ \hline 
\endhead

\hline \multicolumn{6}{|r|}{{Continued on next page}} \\ \hline
\endfoot

\hline \hline
\endlastfoot

  1 & Spin Glasses; Magnetic Frustration & 1558 & Dense GC & 1765 &  Treelike GC \\
  \hline
  2 & Complex Networks; Population Dynamics & 2677 & Dense GC & 731 & No GC \\
  \hline
  3  & Bose-Einstein Condensates & 1020 & Dense GC & 105 & No GC \\
  \hline
  4 & Superconducting Phases; High Pressure Phases & 695 & Dense GC & 3780 & Dense GC \\
  \hline
  5 & Quantum Computing & 1135 & Dense GC & 677 & Dense GC \\
  \hline
  6 & Quantum Oscillators & 523 & No GC & 238 & No GC \\
  \hline
  7 &  Spins in Materials; Spintronics & 840 & Dense GC & 1461 & Dense GC \\
  \hline
  8 &  Review Articles & 369 & No GC & 429 & No GC \\
  \hline
  9 & Polarons & 60 & No GC & 85 & No GC \\
  \hline
  10 & Phase Transitions; Quantum Phase Transitions & 60 & No GC & 52 & No GC \\
  \hline
  11 &  Quantum Dots; Mesoscale Physics & 821 & Dense GC & 6489 & Dense GC \\
  \hline
  12 & Transport Measurements & 366 & Treelike GC & 1189 & Treelike GC \\
  \hline
  13 &  Solitons; Stationary States & 280 & Treelike GC & 233& No GC \\
  \hline
  14 &  Superconductor Vortices & 450 & Dense GC & 756 & Dense GC \\
  \hline
  15 & Microscopy & 122 & No GC & 439 & No GC \\
  \hline
  16 & Electronic Devices & 324 & No GC & 5375 & Dense GC \\
  \hline
  17 & Mathematical Physics & 752 & Dense GC & 786 & Treelike GC \\
  \hline
  18 & Spin Chains \& Lattices & 1539 & Dense GC & 1623 & Dense GC \\
  \hline
  19 &  Magnetic Material Properties & 1087 & Dense GC&  5714 & Dense GC \\
  \hline
  20 &  Ultracold Atoms Dynamics & 1300 & Dense GC & 104 & No GC \\
  \hline
  21 & Oxide Thin Films & 1116 & Treelike GC & 54823 & Dense GC \\
  \hline
  22 & Nuclear Magnetic Resonance & 196 & Treelike GC & 620 & Dense GC \\
  \hline
  23 &  Quantum Hall Effect & 742 & Dense GC & 799 & Dense GC \\
  \hline
  24 & Electronic Devices; Nanoscale Devices & 422 & Treelike GC & 3663 & Dense GC \\
  \hline
  25 &  Soft Condensed Matter; Polymer Physics & 1276 & Treelike GC & 993 & No GC \\
  \hline
  26 & Disordered Systems & 444 & No GC & 285 & No GC\\
  \hline
  27 & Optics; Metamaterials & 676 & Treelike GC & 2125 & Dense GC \\
  \hline
  28 & Glasses & 1187 & Dense GC & 844 & Dense GC \\
  \hline
  29 &  Graphene & 402 & Treelike GC & 392 & Treelike GC \\
  \hline
  30 &  Topological Phases & 913 & Dense GC & 433 & Dense GC \\
  \hline
  31 & Simulation Methods; Monte Carlo & 754 & No GC & 356 & Treelike GC  \\
  \hline
  32 &  Mesoscale Transport & 1416 & Dense GC & 1774 & Dense GC \\
  \hline
  33 &  Inelastic Scattering Experiments & 220 & Treelike GC & 756 & Dense GC \\
  \hline
  34 &  Condensed Matter Theory & 641 & Treelike GC & 100 & No GC \\
  \hline
  35 & Soft Condensed Matter; Granular Physics & 1375 & Dense GC & 421 & No GC \\
  \hline
  36 & Superconducting Devices & 596 & Dense GC & 1191 & Dense GC \\
  \hline
  37 &  Thermodynamics & 1574 & Treelike GC & 140 & No GC \\
  \hline
  38 &  Mott-Hubbard Model & 798 & Treelike GC & 901 & Dense GC  \\
  \hline
  39 & Electronic Spectra; ARPES & 457 & Dense GC & 1451 & Dense GC \\
  \hline
  40 & Critical Phenomena & 786 & Treelike GC & 109 & No GC \\
  \hline
  41 & Mechanical Material Properties & 525 & No GC & 2345 & Dense GC \\
  \hline
  42 & Quantum States & 62 & No GC & 29 & No GC \\
  \hline
  43 & Cuprate Superconductors & 1030 & Dense GC & 858 & Dense GC \\
  \hline
  44 & Ferroelectrics & 1043 & Dense GC & 2591 & Dense GC \\
  \hline
  45 & Condensed Matter Theory & 1595 & Treelike GC & 467 & No GC \\
  \hline
  46 &  Superconductivity & 576 & Dense GC & 528 & Dense GC  \\
  \hline
  47 & Surface Physics; Surface Chemistry & 467 & No GC & 938 & Dense GC \\
  \hline
  48 & Density Functional Theory & 1002 & Treelike GC & 8439 & Dense GC \\
  \hline
  49 &  Nonequlibrium Stat Mech& 993 & Treelike GC & 130  & No GC \\
  \hline
  50 & Soft Condensed Matter; Structured Fluids & 432  & Treelike GC & 139 & No GC \\

  \hline
\end{longtable}
\end{center}

  \section{Author Name Disambiguation}\label{appendix3}
  Determining the identities of document authors in publication data sets like the arXiv and Web of Science is a complex open problem in information science (Cf. \cite{bhattacharya2007collective, song2007efficient}).  The analysis in the main text is conducted by labeling each article's author using their their full names as reported in the metadata.  Punctuation is removed and all letters are set to the same case, such that ``Lindsay M. Barnes'' becomes ``lindsay m barnes'' for the purposes of labeling an author.  We adopt this convention because it avoids artificially combining multiple authors with the same name into a single node.  At the same time, this convention is more likely to split single authors across multiple labels.  For example, ``H. Eugene Stanley,'' ``H. E. Stanley,'' and ``H. Stanley'' would all be treated as separate entities in our co-authorship networks.  

  To test whether the results reported above are robust to the choice of naming convention, we repeat our analysis by naming authors according to their first initials and last names.  With this convention, ``Lindsay M. Barnes'' becomes ``l barnes.''  Examining the distribution of articles produced by each author, the new convention does appear to create a large number of entities with common names (e.g. ``h kim,'' ``y lee'') with a far above average publications count.  These entities appear to be composites of many different authors with similar names.

  We reconstruct and measure the co-authorship networks using both author labeling conventions.  The plots in Figure \ref{disambiguation} show measurements of each co-authorship network, similar to the measurements plotted in Figure \ref{measurements} above.  Measurements of the co-authorship networks generated using the first initials and last names are plotted in red, while the measurements of the co-authorship networks generated using full names are plotted in blue.
  The qualitative behavior of the co-authorship network assembly is largely consistent across all topics, with the full author name co-authorship graphs' giant components tending to be smaller and slower to develop. Despite these quantitative differences, the qualitative behavior seen for each topic's network assembly remains consistent between the two naming conventions.  For Topic 5 and Topic 19, each case's respective co-authorship network forms a giant component whose mean path length grows and then shrinks.  For Topic 12, each case's co-authorship network forms a treelike largest connected component.  For Topic 41, each case's network fails to form a connected component.  

  Overall, the results reported in the main text are robust to changing how author nodes are labeled, but the full name convention is more conservative about artificially creating composite author nodes.

  \begin{figure*}[ht!!]
      \centering
      \vspace*{2mm}
  \includegraphics[width=1.\textwidth]{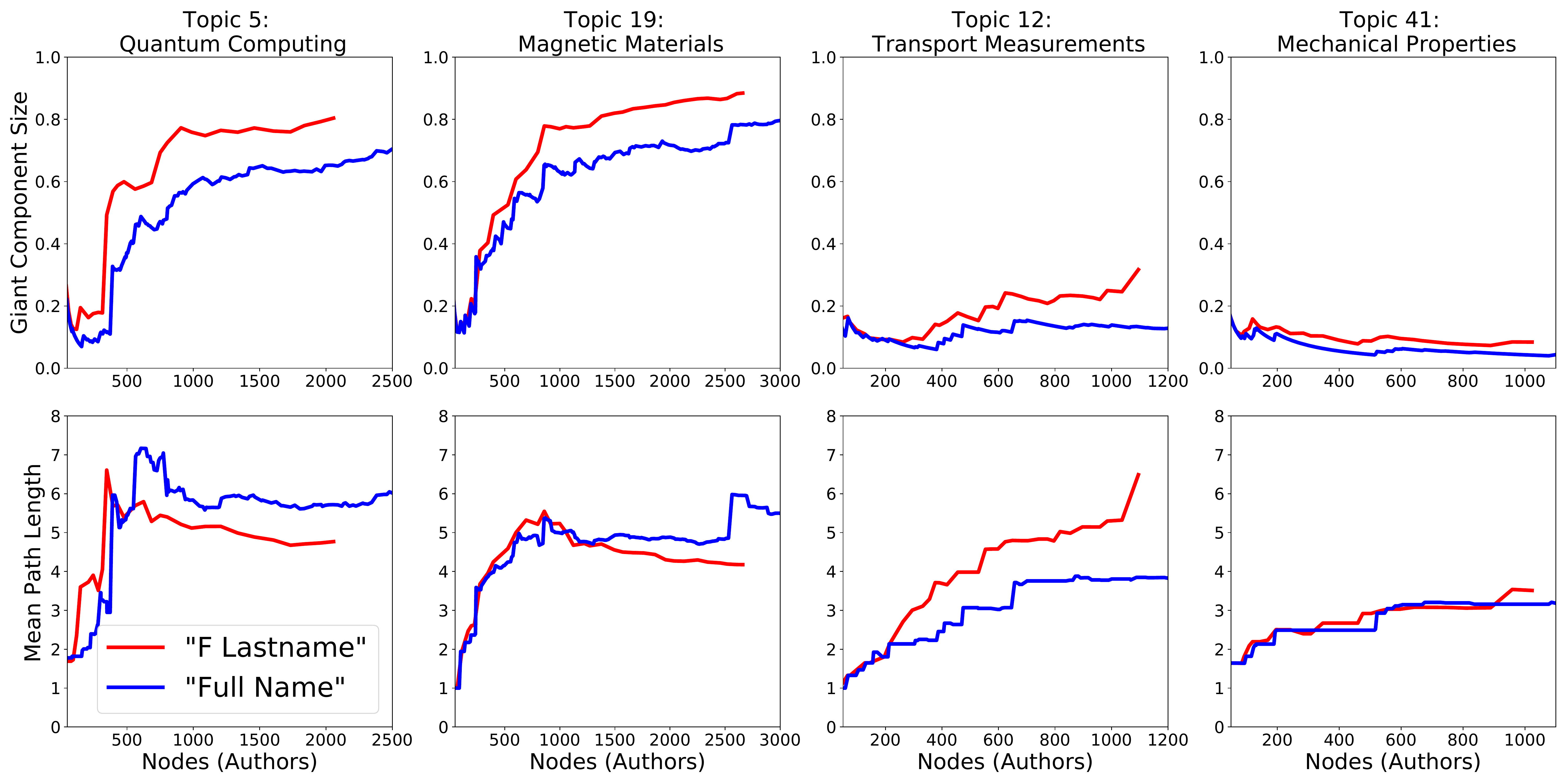}
  \caption{Comparison between co-authorship networks drawn using two different naming conventions.  Each panel shows measurements of co-authorship networks using two different node labeling conventions.  Measurements of the co-authorship networks generated using the first initials and last names are plotted in red, while the measurements of the co-authorship networks generated using full names are plotted in blue. The top row shows the fraction of nodes belonging to the largest component as a measure of network size, plotted vs. the total number of nodes in the network.  The bottom row shows the mean geodesic path length of the largest component (``mean path length'') vs. the total number of nodes in the network.} \label{disambiguation}
  \end{figure*}

  \clearpage

  \section{Modularity}\label{appendix4}

  We evaluate the extent to which topic modeling is effective at detecting distinct communities of authors using a modularity score.  The modularity score, as defined in \cite{newman2010networks} measures the extent to which nodes of a given type (for example, associated with a particular topic) tend to form links with other nodes of the same type vs. nodes of a different type.  

  For a pair of topics $X$ and $Y$, we calculate the modularity by first identifying the authors who have written articles associated with either or both topics.  Based on co-authorship, we then identify all edges between authors belonging to the same topic.  If the two connected authors are both only associated with topic $X$ or only with topic $Y$, then this edge connects like with like.  (We consider cases where an author in topic $X$ connected to an author in both topics $X$ and $Y$ to be ``unlike'' edges.)  After identifying these edges, we can then calculate the standard normalized modularity score

  \begin{equation}\label{mod}
  \frac{Q}{Q_{\text{max}}} = \frac{\sum_{ij} \left(A_{ij} - k_i k_j / 2m \right) \delta\left(c_i,c_j\right)}{2m - \sum_{ij} \left( k_i k_j / 2m \right)\delta\left(c_i,c_j\right) }
  \end{equation}

  where $m$ is the total number of edges, $A$ is the adjacency matrix, $k_i$ is the degree of the $i$th author, $c_i$ is the community that the $i$th author belongs to, and the sum is performed over all authors \cite{newman2010networks}.  A modularity score close to one means that there are relatively few links between nodes of different types.  A modularity score of less than one means that the different types of nodes are mixed.  A network consisting of nodes of only a single type has modularity 0.

  In the case of our co-authorship networks, we label our author nodes according to whether they have written one or more articles that is strongly associated with one of the topics detected using LDA.  The modularity scores for each pair of topics is shown in Figure \ref{modularity}, and a histogram of values is shown in Figure \ref{modu_hist}.  For the majority of pairs of topics, the modularity scores are close to 1, meaning that the communities of authors associated with different topic are only sparsely connected to one another.  There are a few exceptions where the modularity score is lower (in the left tail of Figure \ref{modu_hist}), but these are cases where the topics are closely related to one another and one expects there to be a lot of overlap between the two topics.  For example, Topics 3 and 20 are both related to ultracold atoms, while Topics 43 and 46 are both related to superconductors and superconductivity.

  \begin{figure*}[ht!!]
      \centering
      \vspace*{2mm}
  \includegraphics[width=.4\textwidth]{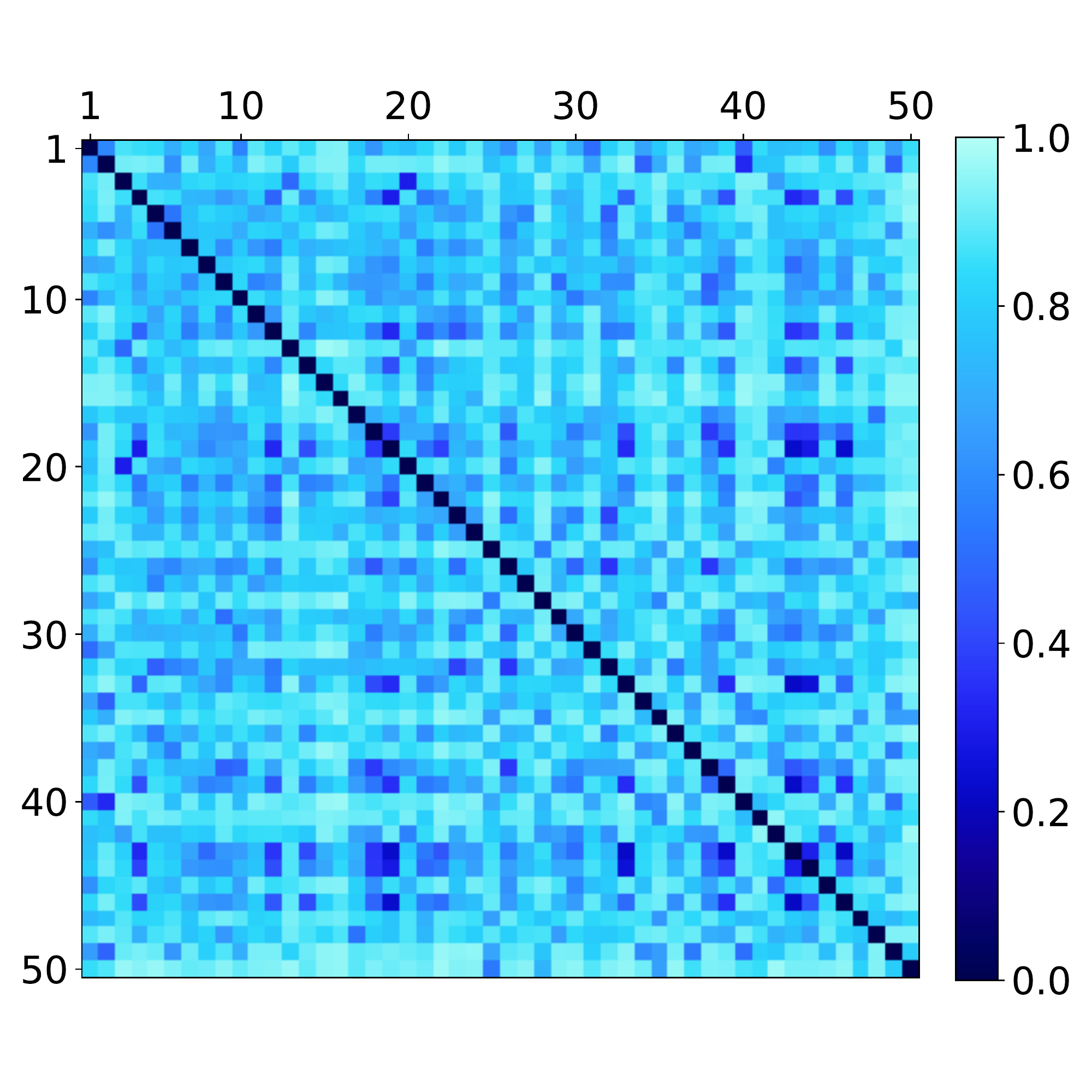}
  \caption{Modularity scores for the co-authorship networks of each pair of topics.
  } \label{modularity}
  \end{figure*}

  \begin{figure*}[ht!!]
      \centering
      \vspace*{2mm}
  \includegraphics[width=.4\textwidth]{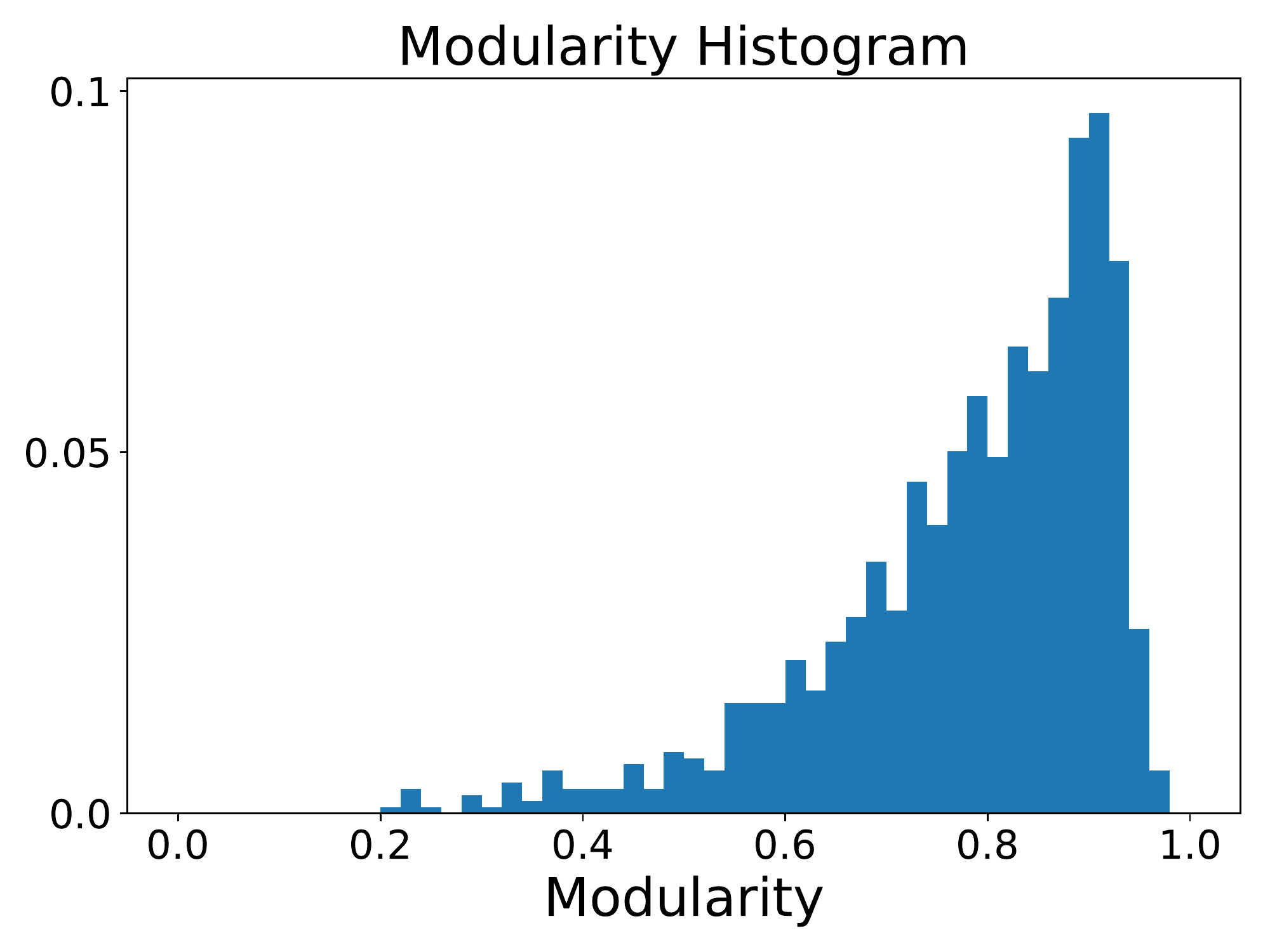}
  \caption{Histogram of modularity scores for all pairs of topics.  A modularity score close to one means that two topics are mostly separated from one another.  The left tail includes pairs of topics that are more closely related to one another, meaning that one expects the two communities to overlap more in those cases.} \label{modu_hist}
  \end{figure*}

  \clearpage

  \section{Robustness to Edge Removal, Visualizations}\label{appendix5}

  Figure \ref{lifetime} in the main text shows how the quantitative measurements of four topics change if an edge is removed from the network if a collaboration is not repeated after a finite amount of time.  Figures \ref{vis_5}-\ref{vis_38} show the network visualizations that accompany those quantitative measurements.  In each figure, each row shows how the co-authorship network evolves over time for a given edge survival lifetime (that is to say, the four rows in each of these diagrams correspond to the four lines in each of the panels in Figure \ref{lifetime}).  Disconnected single nodes are not pictured.

  The co-authorship networks for both Topic 5 (Figure \ref{vis_5}) and Topic 19 (Figure \ref{vis_19}) robustly grow to form large connected components for survival times of 5 years.  By contrast, the co-authorship networks for Topic 38 (Figure \ref{vis_38}) are less robust to edge removal, and only for survival times of longer than 10 years does the giant component appear to form.  The co-authorship network for Topic 12 (Figure \ref{vis_12}) only contains a treelike connected component if edges are allowed to survive indefinitely. Allowing edges to be removed only breaks up this largest component.

  \begin{figure*}[ht!!]
      \centering
      \vspace*{2mm}
  \includegraphics[width=.9\textwidth]{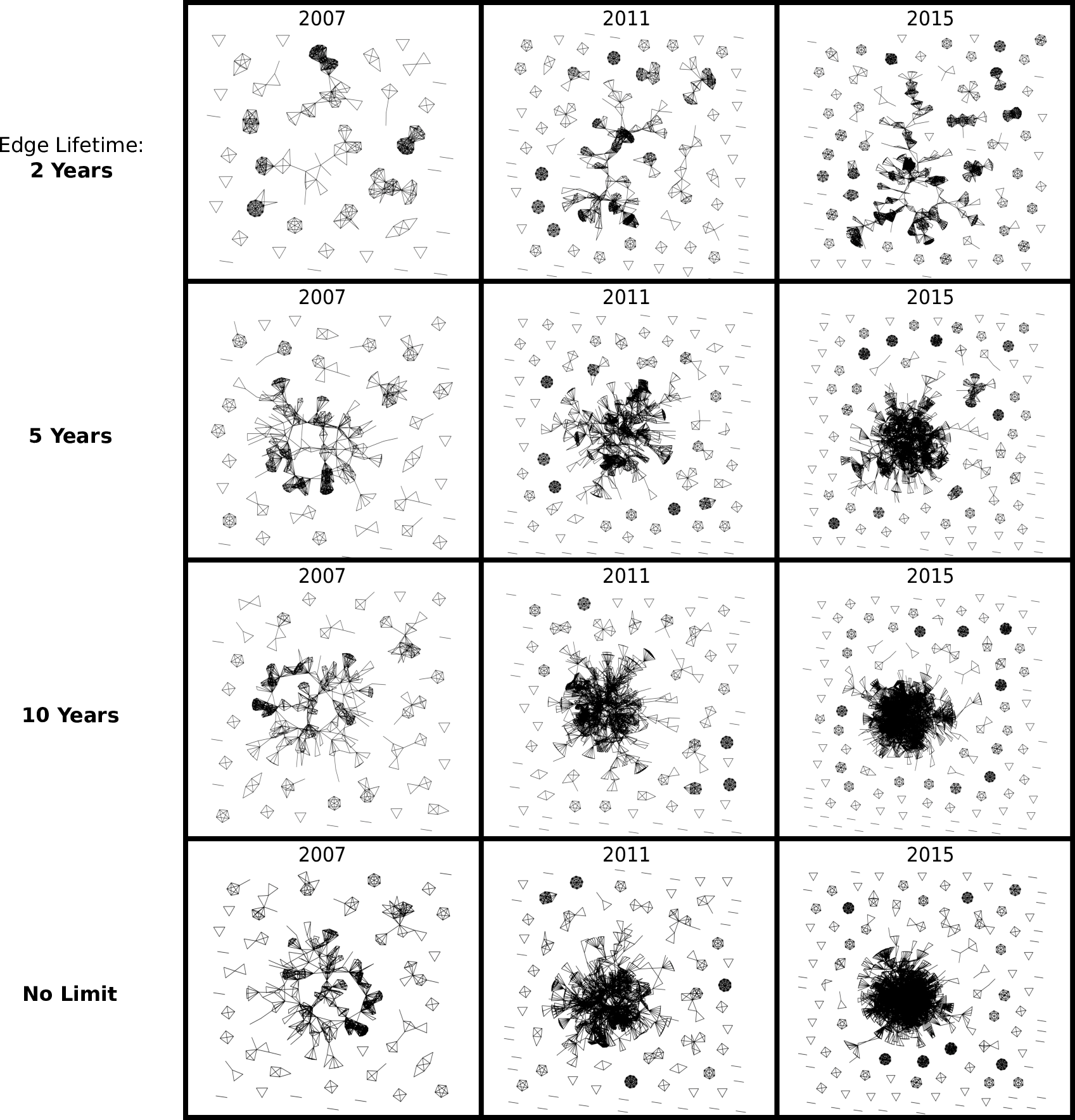}
  \caption{Topic 5 Network Visualizations with Edge Removal: The giant component formation is somewhat suppressed for short survival times (2 years, top row), although a single large connected component does form.  The dense giant component forms robustly for survival times longer than 5 years (bottom three rows). }\label{vis_5} 
  \end{figure*} 

  \begin{figure*}[ht!!]
      \centering
      \vspace*{2mm}
  \includegraphics[width=.9\textwidth]{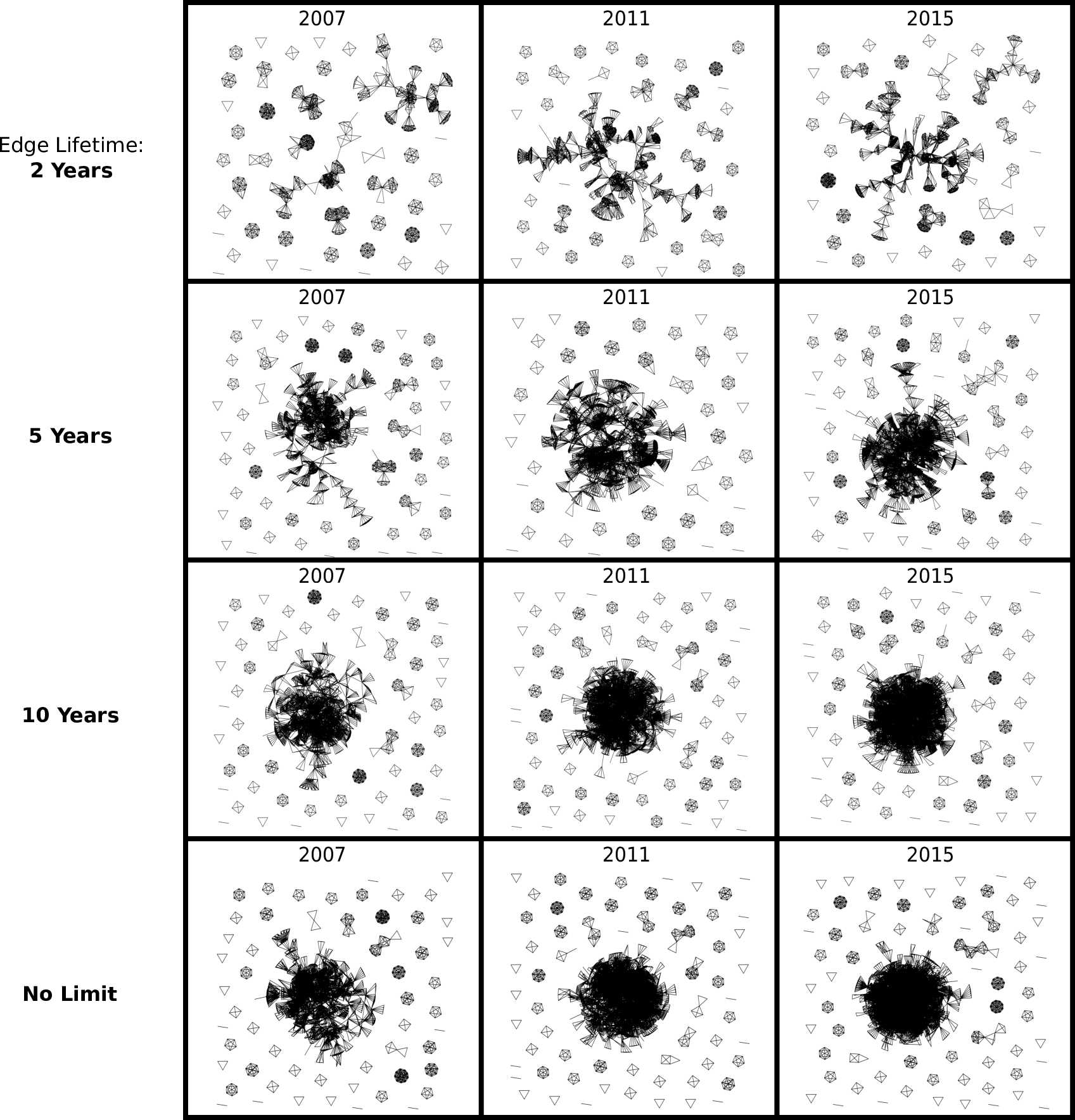}
  \caption{Topic 19 Network Visualizations with Edge Removal: Similar to the network for Topic 5 shown in Figure \ref{vis_5} above, the giant component is somewhat suppressed for short survival times (2 years, top row), although large connected components do form.  The dense giant component forms robustly for survival times longer than 5 years (bottom three rows). }\label{vis_19}
  \end{figure*} 

  \begin{figure*}[ht!!]
      \centering
      \vspace*{2mm}
  \includegraphics[width=.9\textwidth]{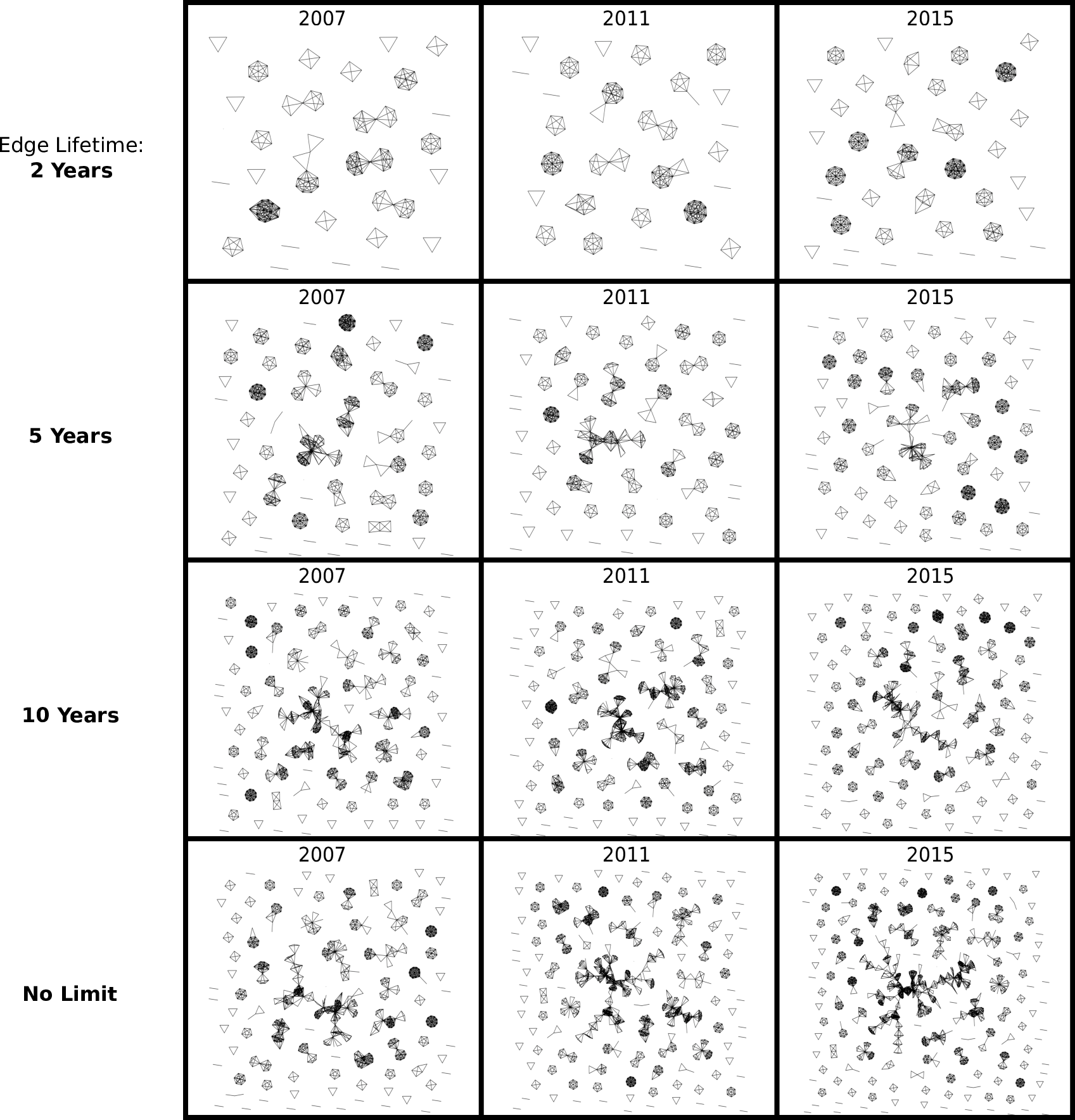}
  \caption{Topic 12 Network Visualizations with Edge Removal: The co-authorship network only forms relatively small treelike connected component if edges are allowed to survive indefinitely. Allowing edges to be removed only breaks up this largest component.}\label{vis_12}
  \end{figure*} 

  \begin{figure*}[ht!!]
      \centering
      \vspace*{2mm}
  \includegraphics[width=.9\textwidth]{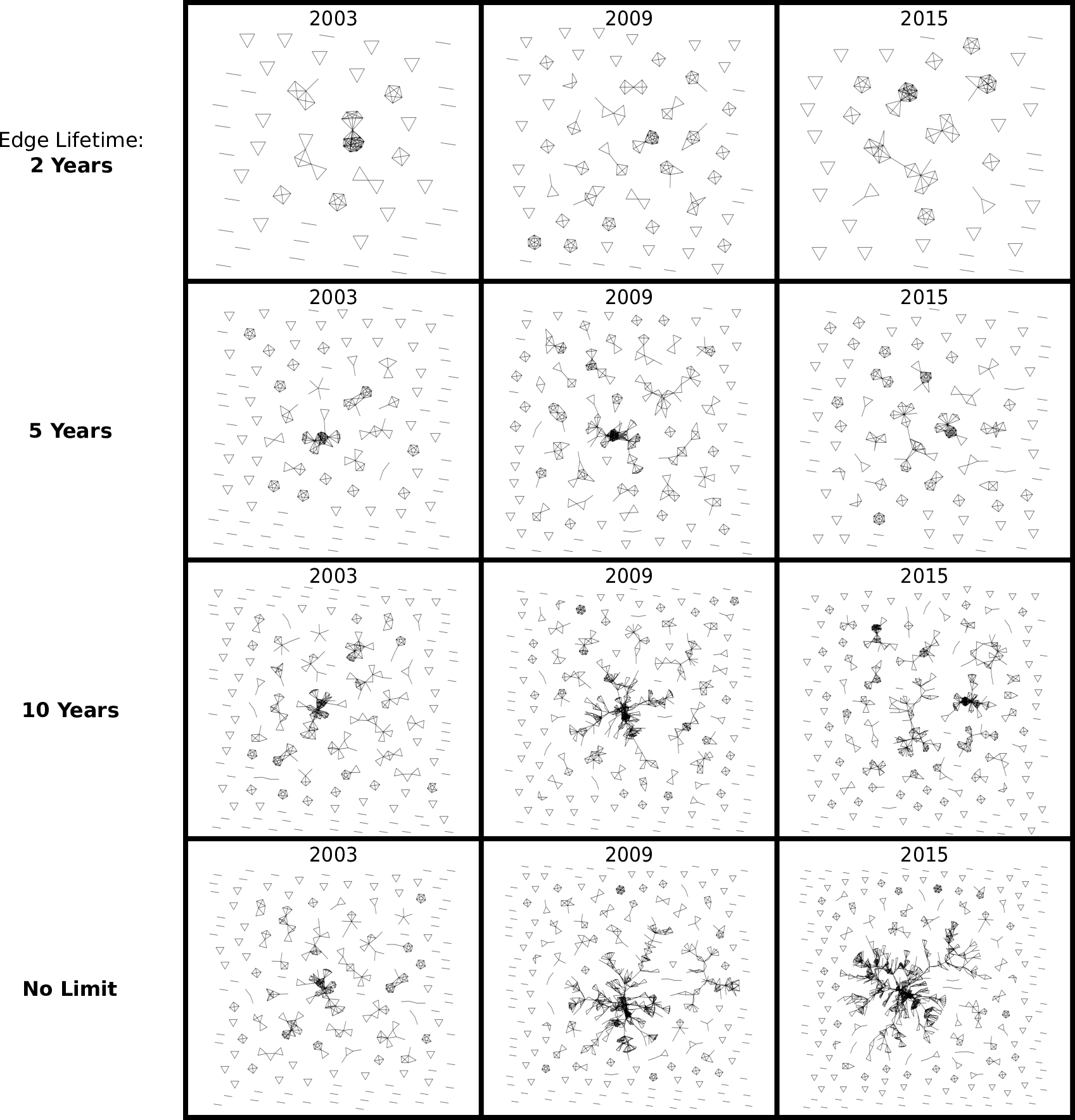}
  \caption{Topic 38 Network Visualizations with Edge Removal: For short survival times (2 years and 5 years, top 2 rows), no component develops.  For longer survival times (10 years, third row), a sparse, treelike large component does form.  Only for very long survival times (bottom row) does the largest component appear to densify.  The co-authorship network for Topic 38 is less robust to removal than those of Topic 5 (Figure \ref{vis_5}) and Topic 19 (Figure \ref{vis_19}).}\label{vis_38}
  \end{figure*} 

  \clearpage

  \section{Review Articles}\label{appendix6}
  The topic model also appears to identify a set of articles as review articles that are distinguished according to their stylistic content.  In addition to reflecting standard research terminology, Topic 8 has keywords such as  ``review,'' ``comment,'' and ``discuss.''  The articles strongly associated with Topic 8 are often also associated with other Topics. 

  Topic 8 is different from a ``junk topic'' in that it contains an interpretable \textit{stylistic} theme without having a coherent \textit{scientific} theme.  The articles associated with Topic 8 have lots of terms in common, but those terms lack the specificity of the scientific terminology detected in some of the other topics.  As such, one cannot expect that the authors who contributed to the articles associated with Topic 8 have common research interests, and so no giant component of collaborators should form in the co-authorship network.  Figures \ref{vis_8} and \ref{measurements_8} show how the cliques in Topic 8's co-authorship network fail to join together.  Even though Topic 8's articles share a common set of terminology, the co-authorship network does not form a giant component.  Only when a topic's keywords reflect a coherent scientific theme does a giant component form.

  \begin{figure*}[ht!!]
      \centering
      \vspace*{2mm}
  \includegraphics[width=.9\textwidth]{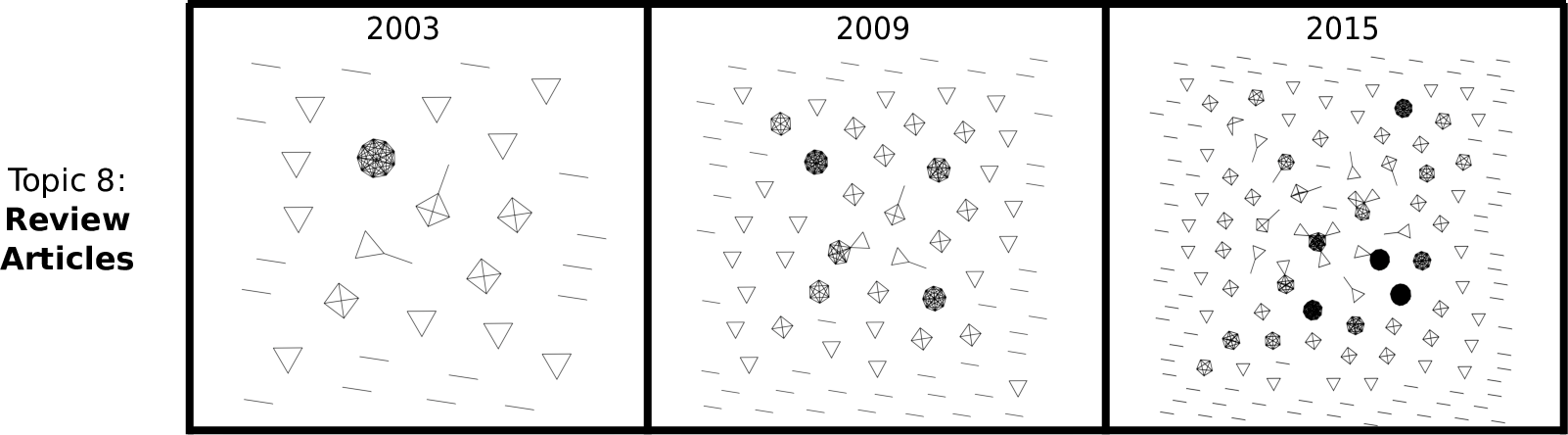}
  \caption{Topic 8 Network Visualizations: The co-authorship network remains a disjointed collection of cliques.}\label{vis_8} 
  \end{figure*} 

  \begin{figure*}[ht!!]
      \centering
      \vspace*{2mm}
  \includegraphics[width=.6\textwidth]{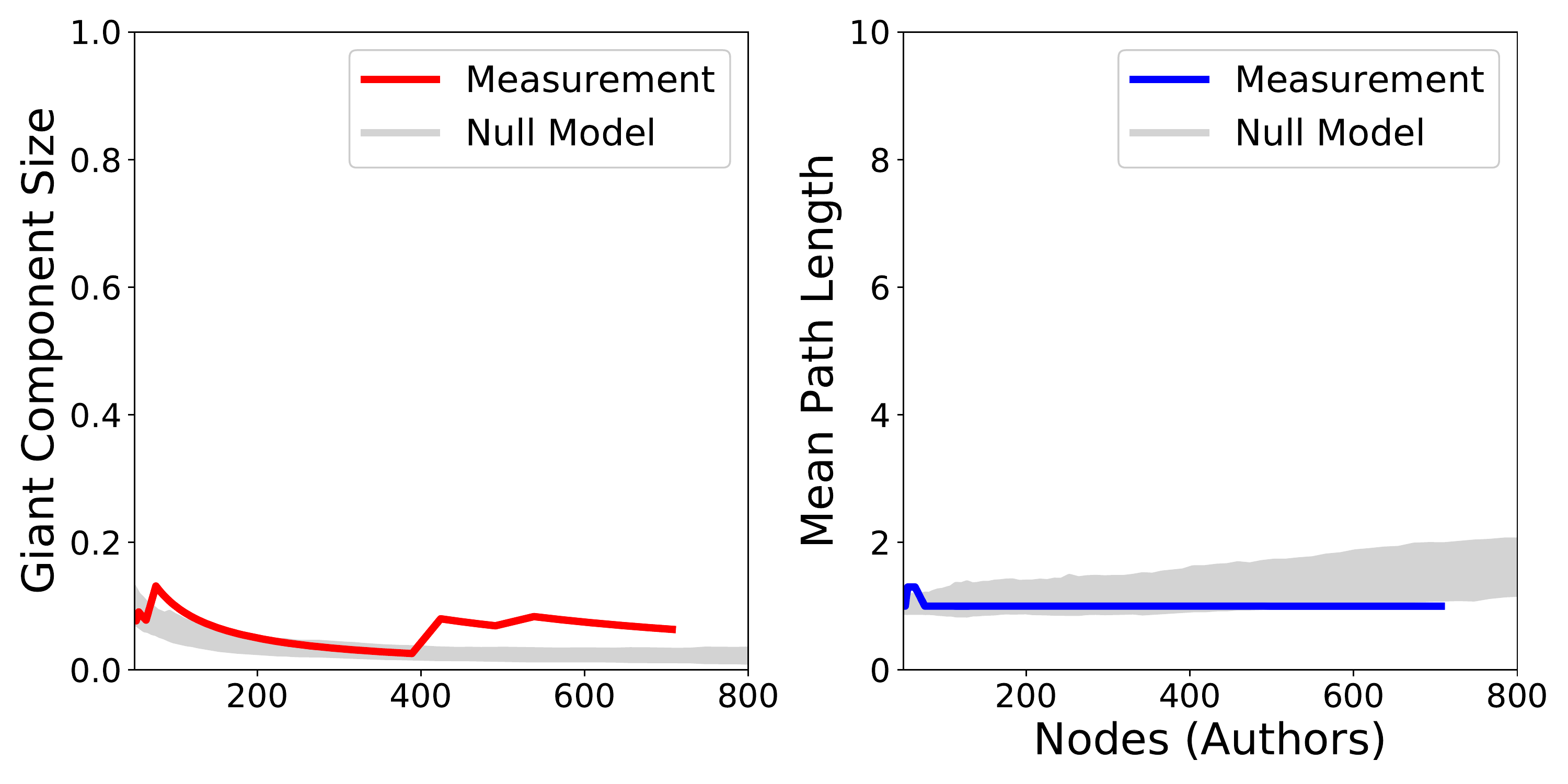}
  \caption{Topic 8 Network Measurements: In contrast to Topic 5 and Topic 19 seen in Figure \ref{measurements}, no giant component forms.}\label{measurements_8} 
  \end{figure*} 

  \end{appendix}
\end{onecolumngrid}

\end{document}